\documentclass[aps,prl,twocolumn,superscriptaddress,showpacs,10pt]{revtex4-1}
\usepackage{graphicx}
\usepackage{braket}
\usepackage{blindtext}
\usepackage{amsmath} 
\usepackage{upgreek}
\usepackage{amssymb}
\usepackage{color}
\usepackage{epstopdf}
\usepackage{tikz}
\usetikzlibrary{decorations.pathmorphing}
\usetikzlibrary{decorations.markings}
\usepackage{overpic}
\usetikzlibrary{patterns}
\usetikzlibrary{shapes}
\usetikzlibrary{arrows}

\begin{document}

\title{Channel Blockade in a Two-Path Triple-Quantum-Dot System}

\author{M. Kotzian}
\email[]{kotzian@nano.uni-hannover.de}
\affiliation{Institut f\"ur Festk\"orperphysik, Leibniz Universit\"at Hannover, Appelstrasse 2, 30167 Hannover, Germany}
\author{F. Gallego-Marcos}
\affiliation{Instituto de Ciencia de Materiales, CSIC, Cantoblanco, 28049 Madrid, Spain}
\affiliation{Institut f\"ur Festk\"orperphysik, Leibniz Universit\"at Hannover, Appelstrasse 2, 30167 Hannover, Germany}
\author{G. Platero}
\affiliation{Instituto de Ciencia de Materiales, CSIC, Cantoblanco, 28049 Madrid, Spain}
\author{R.~J. Haug}
\affiliation{Institut f\"ur Festk\"orperphysik, Leibniz Universit\"at Hannover, Appelstrasse 2, 30167 Hannover, Germany}

\date{\today}

\begin{abstract}
Electronic transport through a two-path triple-quantum-dot system with two source leads and one drain is studied. By separating the conductance of the two double dot paths, we are able to observe double dot and triple dot physics in transport and study the interaction between the paths. We observe channel blockade as a result of inter-channel Coulomb interaction. The experimental results are understood with the help of a theoretical model which calculates the parameters of the system, the stability regions of each state and the full dynamical transport in the triple dot resonances. 
\end{abstract}


\maketitle

\begin{figure}[h]      
       \includegraphics[width=0.48\textwidth]{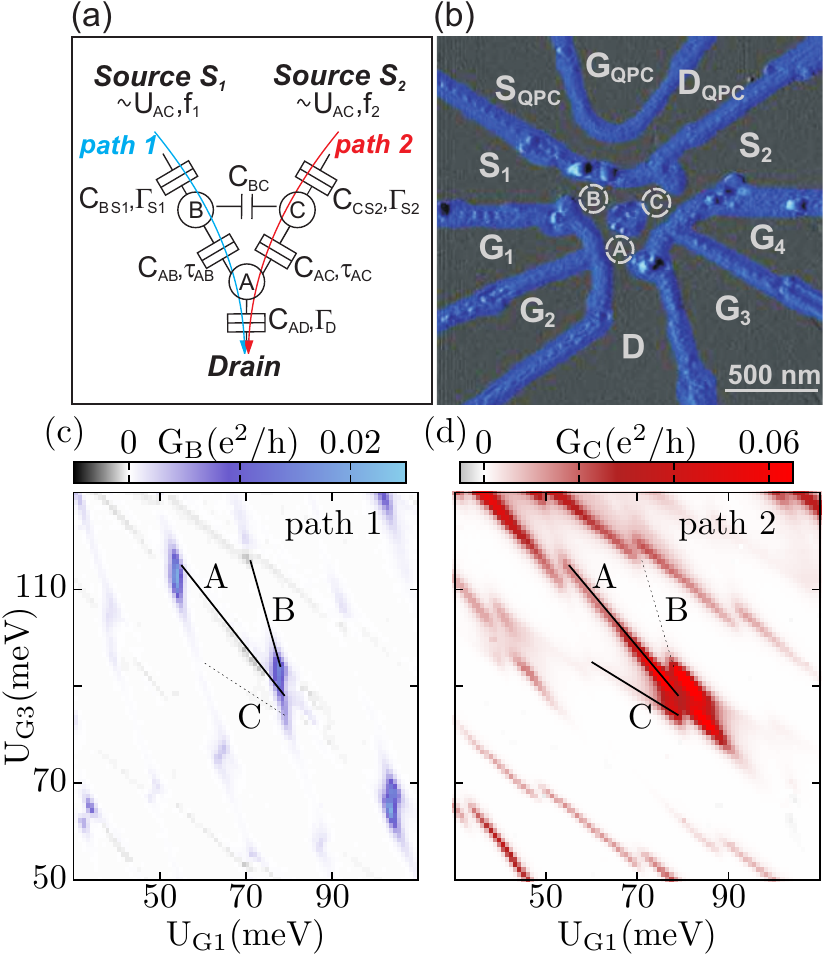}
       \caption{(a) Schematic of the TQD setup with capacitive and tunnel couplings. (b) AFM picture of the TQD sample with the in-plane gates $\text{G}_{1}$ - $\text{G}_{4}$ and a QPC for charge measurements. The blue lines indicate the insulating barriers written by AFM. (c) Transport through path 1. Charging lines of dots A and B (solid lines) are observed. Charging of dot C (dotted line) is observed by a shift of the charging lines. (d) Transport through path 2. Charging lines of dot A and C (solid lines) can be seen and charging of dot B (dotted line) is observed by a shift of these charging lines.
       }\label{fig:sample}
\end{figure}

Triple quantum dots (TQDs), which have been implemented only recently \cite{Vidan2004,Gaudreau2006,Schroeer2007,Rogge2008}, offer the possibility of analyzing new fascinating properties which are not present in double-quantum-dot systems. These new properties, to name a few, include interference phenomena between different transport channels giving rise to dark states in triangular \cite{Michaelis2006,Poeltl2009,Busl2010,Emary2007} and linear \cite{Sanchez2014a} dot distributions and long distant coherent states in TQDs \cite{Sanchez2014a,Gallego-Marcos2015,2013NatNa...8..261B,2013NatNa...8..432B,PhysRevLett.112.176803}. 
TQDs are, as the smallest qubit chain, a step towards more complex architectures needed in quantum computation. They allow for novel applications in the field of quantum information processing, like for example as exchange-controlled spin qubits \cite{2012NatPh...8...54G,DiVincenzo2000} or as current rectifiers \cite{Vidan2004,Stop2002}. They provide as well the implementation of quantum cellular automata processes, a combination of charging and reconfiguration events in the system being a crucial process in quantum information \cite{Lent1993,Rogge2009}. Coherent electron transfer using adiabatic passage was proposed for TQDs in series \cite{Greentree2004}. Furthermore, decoherence due to charge fluctuations is reduced in a TQD-based coded qubit as it involves a decoherence free subspace \cite{DiVincenzo2000,Sasakura2004}.\\
Our system is a triangular-shaped TQD with one lead attached to each dot thus consisting of two double-dot paths. A triangular geometry is suitable for studying entanglement and effects of interference which makes it an interesting device for quantum information processing. The flexibility of this setup makes it a convenient tool for investigating the transport properties of a TQD. 
Transport can be measured separately and simultaneously for the two double dot paths and be compared or combined to study the whole TQDs physics on the basis of the double dots. Also, transitions from double dot resonances in one path to configurations of all three dots in resonance can be studied in transport.  
In contrast to former published works \cite{Rogge2008} where one source and two drain leads were used, we now use one drain and two source leads. In this configuration of two-path transport the dot connected to the drain is shared by both paths (Fig.\ref{fig:sample} (a)). The electrons from the different paths compete for the occupation of this dot. We analyze the role of interactions between the charge flowing through the two different paths by transport measurements. We observe, as a consequence of inter-channel Coulomb interaction, channel blockade in transport.

\textit{TQD Sample and Characterization}. The measurements were performed on a lateral TQD made with local anodic oxidation by atomic force microscope (AFM) on GaAs/AlGaAs-heterostructure \cite{M.Ishii1995,Held1998,Keyser2000}.
A two-dimensional electron gas with an electron concentration of $\text{n}_\text{e}=3,47\cdot10^{15}\text{m}^{-2}$ is located in 33 nm depth below the surface.
The dots A,B,C are arranged in a triangular geometry \cite{Rogge2008} with each dot placed next to the other two and one lead attached to each dot (Fig.\ref{fig:sample} (a)). Dots A and B and also A and C are tunnel coupled, dots B and C are capacitively coupled only.
The source leads $\text{S}_{1}$ and $\text{S}_{2}$ are connected to dots B and C respectively and dot A is connected to the drain lead D. We have two transport paths: path 1 with dots A and B and path 2 with dots A and C.
The sample has four in-plane gates $\text{G}_{1}$-$\text{G}_{4}$ (Fig.\ref{fig:sample} (b)) to control the potential of the dots, interdot and dot-lead couplings. A quantum point contact (QPC) sensitive to all three dots is placed next to dots B and C to perform charge measurements.
The measurements were conducted in a dilution refrigerator. 
To measure the differential conductance of the two transport paths simultaneously but separately, a lock-in technique was used with ac voltages with two different frequencies $\text{f}_{1}$ = 83.3 \text{Hz} and $\text{f}_{2}$ = 18.3 Hz with $\text{U}_{\text{AC}}$ = 10 $\upmu$V applied to $\text{S}_{1}$ and $\text{S}_{2}$ respectively.
The QPC was operated by applying a dc voltage to the source of the QPC, $\text{S}_{\text{QPC}}$, and measuring a dc current at the drain of the QPC, $\text{D}_{\text{QPC}}$. The QPC is tuned by the gate $\text{G}_{\text{QPC}}$.
In our transport measurement range, the dots contain several ten electrons on the whole. The charging energies are $\text{E}_{\text{ch,A}}$ = 2 meV, $\text{E}_{\text{ch,B}}$ = 6 meV, $\text{E}_{\text{ch,C}}$ = 3 meV for dot A, B and C respectively.
\\

\begin{figure}[t]
\includegraphics[width=0.48\textwidth]{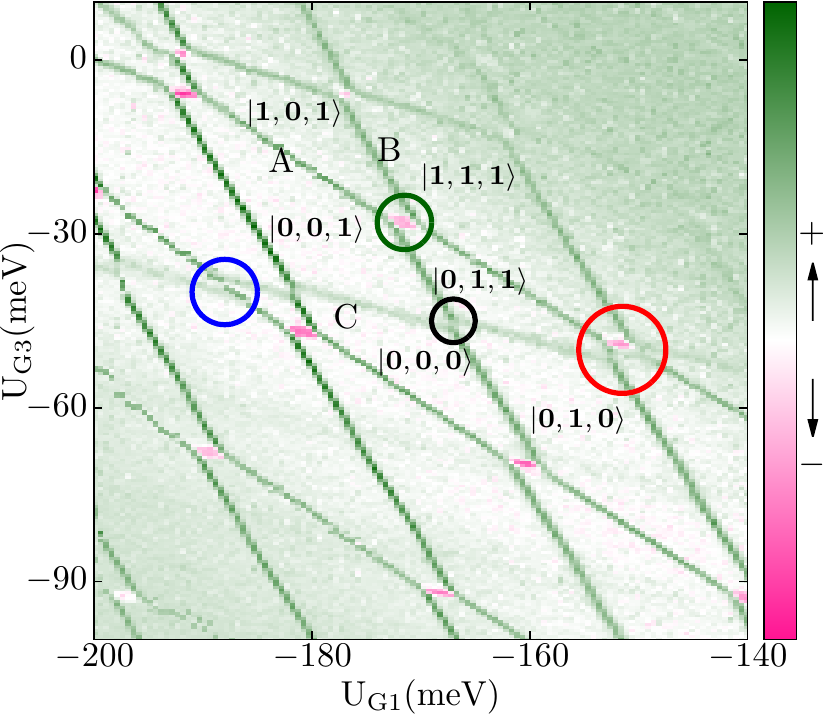}%
\caption{Charge measurement using the QPC. The green lines with different slopes point out the charging of the respective dot with one electron. Double dot resonances are marked by circles in green, black and blue, the triple dot resonance by a red circle.}\label{fig:qpc}
\end{figure}

\textit{Charge Measurements.} To characterize the device, the charging is studied by using the QPC as a detector. The derivative of the QPC current is plotted as a function of gate voltages $\text{U}_{\text{G1}}$ and $\text{U}_{\text{G3}}$ (Fig.\ref{fig:qpc}) with denoted charge configurations $\ket{\text{N}_{\text{A}},\text{N}_{\text{B}},\text{N}_{\text{C}}}$, where $\text{N}_{\text{i}}$ are the occupations of dots A,B,C. The electrons in the core of the dots are not included in $\text{N}_{\text{i}}$. The green lines indicate charging events, where one more electron is added to the system, pink lines indicate electron movement away from the detector. Charging lines with three different slopes, one slope for each dot, are visible in the measurement, as the slope depends on the capacitive coupling and thus on the distance between the dot and gates $\text{G}_{1}$ and $\text{G}_{3}$. The lines with the lowest slope belong to dot C as it is the least coupled to $\text{U}_\text{G1}$, the lines with intermediate slope to dot A and the lines with the largest slope to dot B as it is the least coupled to $\text{U}_{\text{G3}}$. Anticrossings of two dots in resonance are visible where two charging lines meet. At such a resonance two triple points (TPs) emerge where at each one three different charge configurations are degenerate. Charge reconfiguration lines connecting the TPs mark the charge transitions between the dots. Resonances between dot A and B (green circle), A and C (blue circle) and also between the only capacitively coupled dots B and C located in the two different paths (black circle) are observed \cite{Elzerman2003}. 
When the double dot anticrossings coincide, the resonance condition for all three dots is fulfilled (red circle). We will focus on such a region in the transport measurements.\\

\begin{figure}[t]
\includegraphics[width=0.48\textwidth]{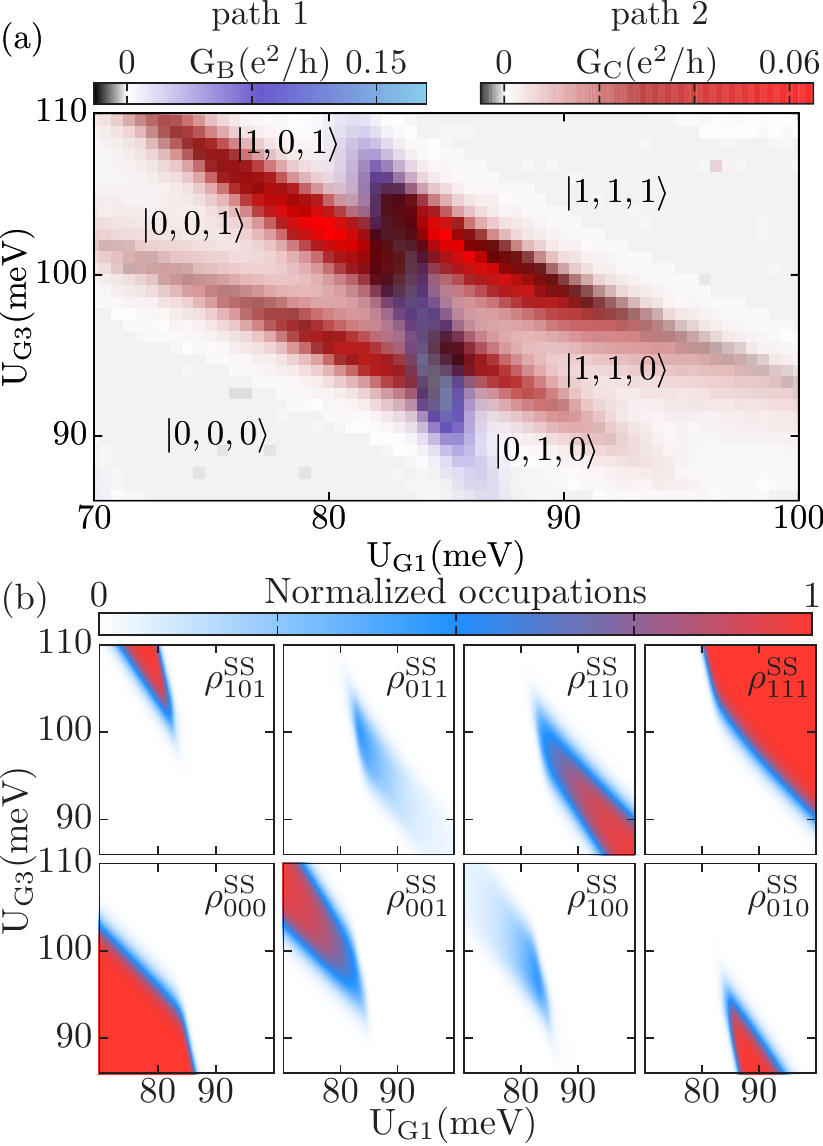}\\
\caption{(a) Combined color plot of differential conductance through path 1 and 2 with denoted charge configurations of the stability regions. (b) Calculated steady state occupation probabilities $\rho_{i}^{SS}$ of each state $i$ in the stability regions of the system (bottom). States $\left|1,0,0\right\rangle$ and $\left|0,1,1\right\rangle$ are a small region with low occupation probability.
}\label{fig:resonance}
\end{figure}

\begin{figure}[t]
\includegraphics[width=0.48\textwidth]{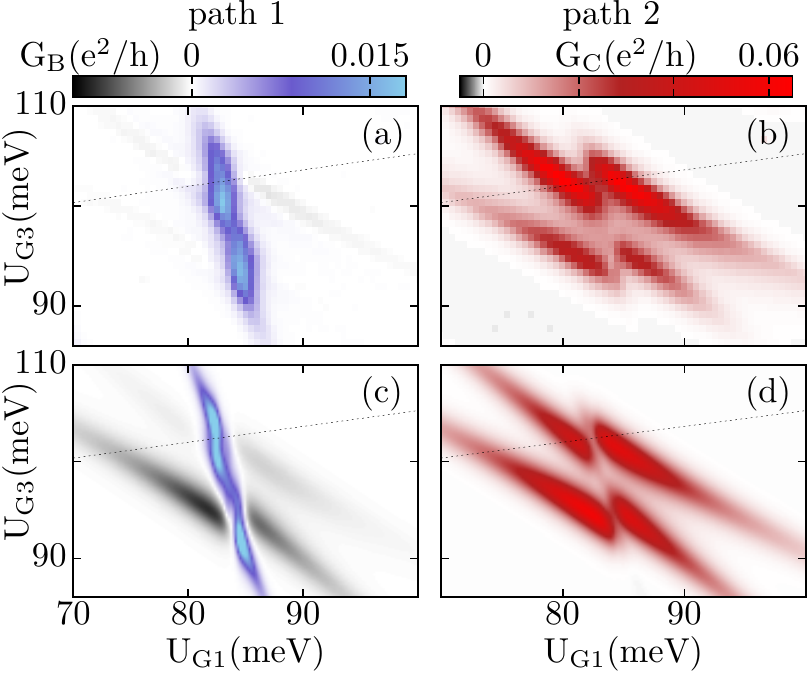}%
\caption{Differential conductance in experiment along path 1 (a) and path 2 (b) and in simulation along path 1 (c) and path 2 (d). The dotted line is the cut plotted in Fig.\ref{fig:cut}.}\label{fig:blockade}
\end{figure}

\textit{Transport Measurements.} To understand transport in this system, the differential conductance G is measured along path 1 and 2 simultaneously but separately, sweeping gate voltages $\text{U}_{G1}$ and $\text{U}_{G3}$ (Fig.\ref{fig:sample} (c),(d)). In doing so, the QPC is not in use. TPs with finite differential conductance are observed in both paths where the two dots are in resonance. In path 1 (Fig.\ref{fig:sample} (c)) resonances of the dots A and B can be seen. Charging of dot C is observed by a shift of the charging lines of dot A and B, where dot C comes into resonance. Analogously, charging lines of dot A and C appear in path 2 (Fig.\ref{fig:sample} (d)) and charging of dot B is detected indirectly by the shift of the charging lines of dot A and C.
\\
A triple dot resonance is formed where two double dot resonances coincide. 
In Fig.\ref{fig:resonance} (a) where we combine path 1 and path 2 as observed in Fig.\ref{fig:sample} (c),(d) we have three double dot resonances, A and B, A and C, B and C, in close vicinity to each other. One observes regions of high differential conductance in both paths but at different gate voltages $\text{U}_{\text{G1}}$ and $\text{U}_{\text{G3}}$. Along the B charging line (path 1) we can identify two resonance lines where A is resonant with B. The whole triple dot physics can be observed.
The different occupations of the states at the TP participating in transport are marked in Fig.\ref{fig:resonance}. 
\\
To understand the experimental results in more detail a theoretical model is developed which reproduces and explains the transport properties of the system. We fit the transport simulations to the experimental data to extract the interdot tunnel couplings $\tau_{\text{AB}}$ = 0.012 meV, $\tau_{\text{AC}}$ = 0.020 meV, the dot-lead tunnel couplings $\Gamma_{\text{D}}$ = 0.008 meV, $\Gamma_{\text{S}_1}$ = 0.003, $\Gamma_{\text{S}_2}$ = 0.006 meV and the electron temperature $\text{T}_{\text{el}}$ = 300 mK. In the model we distinguish between particles coming from $\text{S}_1$ and $\text{S}_2$ in order to be as close as possible to the experimental conditions.
\\
\ 
\\
\textit{Simulation}. Using Master equation techniques (see Supplementary for more detail), we calculate the transport through the two paths and the stability regions of each state. 
\\
In Fig.\ref{fig:resonance} (b) we plot the numerical result of the steady state occupations $\rho_i^\text{SS}$. For small and large $\text{U}_{\text{G1}}$ and $\text{U}_{\text{G3}}$ the states $\ket{0,0,0}$ and $\ket{1,1,1}$ are occupied respectively. Above $\ket{0,0,0}$, for larger values of $\text{U}_\text{G3}$, the state $\ket{0,0,1}$ with one more electron in C becomes occupied and at the right hand side, for larger values of $\text{U}_\text{G1}$, the state $\ket{0,1,0}$ with one more electron in B becomes occupied. Below $\ket{1,1,1}$, for smaller values of $\text{U}_\text{G3}$, the occupation of $\ket{1,1,0}$ with one less electron in C increases and at the left hand side, for smaller values of $\text{U}_\text{G1}$, the state $\ket{1,0,1}$ with one less electron in B is occupied. All these regions obtained numerically correspond perfectly to the ones in Fig.\ref{fig:resonance} (a). The small regions of $\ket{1,0,0}$ and $\ket{0,1,1}$ connect the states with one electron and two electrons respectively. Each of these small regions, not really seen in Fig.\ref{fig:resonance} (a), contain two TPs of path 1 and two TPs of path 2. When two TPs coincide we have a quadruple point.
\\
With the information from the theoretical calculation of the occupations we determine the TPs present on the resonant lines in path 1 (Fig.\ref{fig:resonance} (a)), ($\left|0,0,0\right\rangle$, $\left|0,1,0\right\rangle$, $\left|1,0,0\right\rangle$), ($\left|1,1,0\right\rangle$, $\left|1,0,0\right\rangle$, $\left|0,1,0\right\rangle$) and ($\left|0,0,1\right\rangle$, $\left|0,1,1\right\rangle$, $\left|1,0,1\right\rangle$), ($\left|1,1,1\right\rangle$, $\left|1,0,1\right\rangle$, $\left|0,1,1\right\rangle$) where in the two last TP there is one more electron in C. In near vicinity of these TP there is high positive differential conductance in path 1 due to temperature broadening of the states. Thus they merge and form a vertical line of high differential conductance. Similarly, for path 2 we have high positive differential conductance at the TPs ($\left|0,0,0\right\rangle$, $\left|0,0,1\right\rangle$, $\left|1,0,0\right\rangle$), ($\left|1,0,1\right\rangle$, $\left|0,0,1\right\rangle$, $\left|1,0,0\right\rangle$) and, with one more electron in dot B, ($\left|0,1,0\right\rangle$, $\left|0,1,1\right\rangle$, $\left|1,1,0\right\rangle$), ($\left|1,1,1\right\rangle$, $\left|0,1,1\right\rangle$, $\left|1,1,0\right\rangle$).
\\
In Fig.\ref{fig:blockade} we plot the measured conductance of path 1 and 2 separately ((a),(b)) as well as the results from the simulation ((c),(d)). In path 1(2) we observe the splitting of the resonance between the dots A,B(A,C) due to the interaction with dot C(B). We also observe negative differential conductance in path 1 where path 2 has high conductance (grey color in Fig.\ref{fig:blockade} (a),(c)). In the following we will analyze and compare the transport features of the two paths in more detail.
\\
\ 
\\
\textit{Channel Blockade.}
In Fig.\ref{fig:cut} we show a cut from Fig.\ref{fig:blockade}. We observe that the resonance of path 2 splits into two ($\blacktriangle$,$\blacklozenge$) due to the interaction with the third dot present in path 1. Path 1 gets in resonance at $\blacktriangledown$ and partially blocks the other path decreasing its conductance. This point is a quadruple point, four states of the two paths are coexisting in the same region of the stability diagram. The transport through path 2 is stronger than through path 1 ($\tau_{\text{AC}}>\tau_{\text{AB}}$) thus when path 2 is in resonance ($\blacktriangle$,$\blacklozenge$) it totally blocks path 1 decreasing its conductance even to negative values (more appreciable in Fig.\ref{fig:blockade} (a) and \ref{fig:blockade} (c)).
\\
This channel blockade is a consequence of Coulomb interaction between the charges flowing through the two transport channels which share dot A. When the bias voltage increases, the path with higher conductance increases its occupation in dot A blocking access to dot A from the other path and thus decreasing its transport and obtaining negative differential conductance.   
\\
In Fig.\ref{Fig:esquema} we identify the dominating and the blocked transport channels for each resonance ($\blacktriangle$,$\blacktriangledown$,$\blacklozenge$) with the information from the simulation. We show the initial and final states connected by two different transport paths, where in ($\blacktriangle$,$\blacklozenge$) one path blocks the other and in ($\blacktriangledown$) both paths share the conductance.
\\
Blockade phenomena were previously studied for one dot attached to three leads (one drain and two sources) which contain some amount of up and down spins \cite{PhysRevB.70.115315}. The spins of each path compete to occupy the dot blocking the access to the dot for the other spin by Coulomb interaction. Compared to this work for a single dot our TQD includes coherences between states of the two paths. Other papers \cite{PhysRevB.89.205424,PhysRevB.85.245431,PhysRevB.82.041303} treat two transport paths just capacitively coupled where at some situations one of the paths blocks the transport through the other by Coulomb interaction.
\\
\begin{figure}[t]
\includegraphics[width=0.48\textwidth]{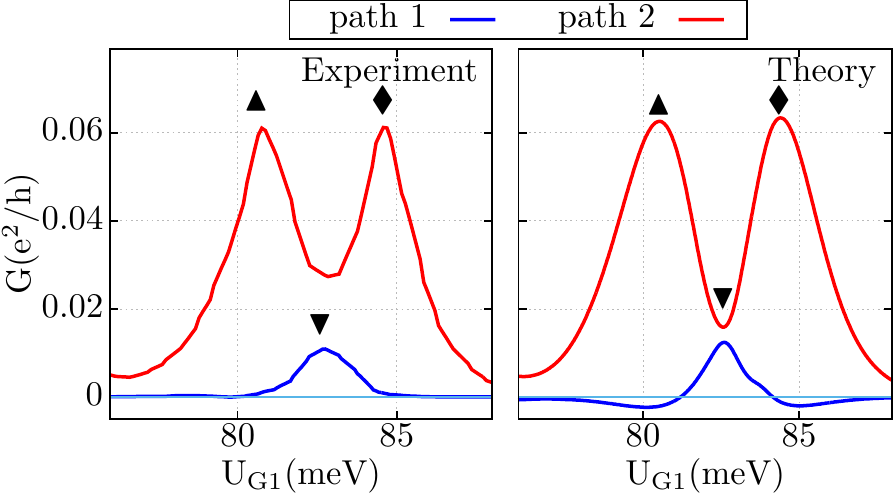}%
\caption{Cut through the transport measurement and simulation of Fig.\ref{fig:blockade} for path 1 and 2 at $\text{U}_{\text{G3}}$=103 mV.}\label{fig:cut}
\end{figure}
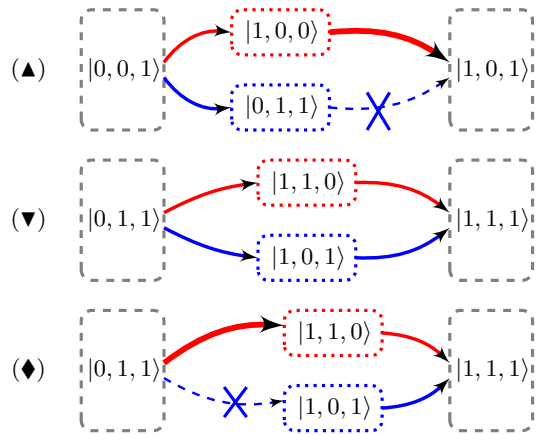
\begin{figure}[h]
\centering
\begin{tikzpicture}[
	>=latex',
	levelL/.style={red,thick},
	levelC/.style={black,thick},
	levelR/.style={orange,thick},
	reservorio/.style={rectangle,thick},
 	trans/.style={<->, bend left},
 	State1/.style={rectangle, rounded corners, fill=white, draw=blue},
 	State2/.style={rectangle, rounded corners, fill=white, draw=red}]
 	MixState/.style={ rounded corners, fill=white, draw=red}]
	
	\draw(-4.2,-0.2) node {($\blacktriangle$)};
	\draw[rounded corners, fill=white, draw=gray!,line width=1.2pt, dashed] (-3.5,-1.0) rectangle (-2.4,0.6) node [below=8mm, left = -1 mm] {$\ket{0,0,1}$};
	\draw[rounded corners, fill=white, draw=red!,line width=1.2pt,dotted] (-1.5,0.0) rectangle (-0.2,0.6) node [below=3mm, left = 0 mm] {$\ket{1,0,0}$};
	\draw[rounded corners, fill=white, draw=blue!,line width=1.2pt,dotted] (-1.5,-1.0) rectangle (-0.2,-0.4) node [below=3mm, left = 0 mm] {$\ket{0,1,1}$};
	\draw[rounded corners, fill=white, draw=gray!,line width=1.2pt, dashed] (1.4,-1.0) rectangle (2.5,0.6) node [below=8mm, left = -1 mm] {$\ket{1,0,1}$};
	
	\draw (-2.4,-0.1) edge[->, bend left=30, draw=red, line width=1.3pt] (-1.5,0.3) node[above=6mm, right = 9 mm] {};	
	\draw (-2.4,-0.3) edge[->, bend right=30, draw=blue, line width=1.3pt] (-1.5,-0.7) node[above=6mm, right = 9 mm] {};	
	\draw (-0.2,0.3) edge[->, bend left=20, draw=red, line width=2.2pt] (1.4,-0.1) node[above=6mm, right = 9 mm] {};
	\draw (-0.2,-0.7) edge[->, bend right=20, draw=blue, line width=0.9pt, dashed] (1.4,-0.3) node[above=6mm, right = 9 mm] {};
	\draw (0.3,-1.0) edge[-, draw=blue, line width=1.2pt] (0.6,-0.5) node {};
	\draw (0.6,-1.0) edge[-, draw=blue, line width=1.2pt] (0.3,-0.5) node {};

	
	\draw[rounded corners, fill=white, draw=red!,line width=1.2pt,dotted] (-1.15,-2.0) rectangle (0.15,-1.4) node [below=3mm, left = 0 mm] {$\ket{1,1,0}$};
	\draw[rounded corners, fill=white, draw=blue!,line width=1.2pt,dotted] (-1.15,-3.0) rectangle (0.15,-2.4) node [below=3mm, left = 0 mm] {$\ket{1,0,1}$};
	\draw[rounded corners, fill=white, draw=gray!,line width=1.2pt, dashed] (-3.5,-3.0) rectangle (-2.4,-1.4) node [below=8mm, left = -1 mm] {$\ket{0,1,1}$};
	\draw[rounded corners, fill=white, draw=gray!,line width=1.2pt, dashed] (1.4,-3.0) rectangle (2.5,-1.4) node [below=8mm, left = -1 mm] {$\ket{1,1,1}$};
	\draw(-4.2,-2.2) node {($\blacktriangledown$)};
		
	\draw (-2.4,-2.1) edge[->, bend left=10,draw=red, line width=1.3pt] (-1.15,-1.7) node[above=6mm, right = 9 mm] {};	
	\draw (-2.4,-2.3) edge[->, bend right=10,draw=blue, line width=1.3pt] (-1.15,-2.7) node[above=6mm, right = 9 mm] {};
	\draw (0.15,-1.7) edge[->, bend left=20, draw=red, line width=1.3pt] (1.4,-2.1) node[above=6mm, right = 9 mm] {};	
	\draw (0.15,-2.7) edge[->, bend right=20, draw=blue, line width=1.3pt] (1.4,-2.3) node[above=6mm, right = 9 mm] {};
	
	
	\draw[rounded corners, fill=white, draw=red!,line width=1.2pt,dotted] (-0.8,-4.0) rectangle (0.5,-3.4) node [below=3mm, left = 0 mm] {$\ket{1,1,0}$};
	\draw[rounded corners, fill=white, draw=blue!,line width=1.2pt,dotted] (-0.8,-5.0) rectangle (0.5,-4.4) node [below=3mm, left = 0 mm] {$\ket{1,0,1}$};
	\draw[rounded corners, fill=white, draw=gray!,line width=1.2pt, dashed] (-3.5,-5.0) rectangle (-2.4,-3.4) node [below=8mm, left = -1 mm] {$\ket{0,1,1}$};
	\draw[rounded corners, fill=white, draw=gray!,line width=1.2pt, dashed] (1.4,-5.0) rectangle (2.5,-3.4) node [below=8mm, left = -1 mm] {$\ket{1,1,1}$};
	\draw(-4.2,-4.2) node {($\blacklozenge$)};
	
	\draw (-2.4,-4.1) edge[->, bend left=20, draw=red, line width=2.2pt] (-0.8,-3.6) node[above=6mm, right = 9 mm] {};	
	\draw (-2.4,-4.3) edge[->, bend right=20, draw=blue, line width=0.9pt, dashed] (-0.8,-4.6) node[above=6mm, right = 9 mm] {};
	\draw (-1.6,-4.42) edge[-, draw=blue, line width=1.2pt] (-1.3,-4.82) node {};
	\draw (-1.3,-4.42) edge[-, draw=blue, line width=1.2pt] (-1.6,-4.82) node {};
	\draw (0.5,-3.7) edge[->, bend left=20, draw=red, line width=1.3pt] (1.4,-4.1) node[above=6mm, right = 9 mm] {};	
	\draw (0.5,-4.7) edge[->, bend right=20, draw=blue, line width=1.3pt] (1.4,-4.3) node[above=6mm, right = 9 mm] {};
	

\end{tikzpicture}
\caption{Transport mechanism for the peaks in Fig.\ref{fig:cut}. In ($\blacktriangle,\blacklozenge$) path 2 blocks path 1 and in ($\blacktriangledown$) both paths share the occupation of dot A.}
\label{Fig:esquema}
\end{figure}
\\
To observe the channel blockade in our experiment, it is crucial to have two sources. If we switch the transport direction of both paths 
 with the source lead connected to dot A and the two drain leads connected to dots B and C, the transport paths influence each other in a different way in the simulations. The electron flow splits at dot A in two paths with a probability that depends on the tunneling rates of path 1 ($\tau_\text{AB},\ \Gamma_\text{B}$) and path 2 ($\tau_\text{AC},\ \Gamma_\text{C}$).
\\
\ 
\\
\textit{Conclusion.} In summary, we have shown channel blockade in electronic transport through a TQD with two source leads. Coulomb interaction between electrons coming from the two sources gives rise to a blockade of transport through one path, when the other path has high conductance. The results show how interaction between the joint transport paths is affecting the transport properties of the multi-terminal device and are a step towards a better understanding of transport properties in complex multi-dot systems.
\\
\begin{acknowledgments} 
We are grateful to M.~C. Rogge for producing the TQD sample. We acknowledge discussions with R. S\'{a}nchez and financial support from Spanish MICINN MAT2014-58241-P.
\end{acknowledgments}

\widetext
\begin{center}
\textbf{\large Supplemental Material: Theoretical Model}
\end{center}
Here we discuss in detail the theory used to simulate the experiment. The total Hamiltonian of the system reads $\hat{H}=\hat{H}_{0}+\hat{H}_{\text{lead}}+\hat{H}_{\text{int}}$. The dot system $(\hat{H}_0)$ is a three site Anderson-like Hamiltonian \cite{1961PhRv..124...41A}: $\hat{H}_{0}=\sum_{i}\epsilon_i\hat{c}^\dag_i\hat{c}_i +\sum_i\tau_{i,i+1}\hat{c}^\dag_i\hat{c}_{i+1}+\sum_{i<j}V_{ij}\hat{n}_i\hat{n}_j$, where $\hat{c}^\dag_i$  is the electron creation operator and $\hat{n}_i$ the particle number operator of dot $i$. $\epsilon_i$ ($i=\{\text{A, B, C}\}$) is the chemical potential of the dots, $\tau_{ij}$ the coherent interdot tunnel coupling and $V_{ij}$ the Coulomb interaction between the electrons in different dots.  The reservoirs are modeled as a Fermi electron gas
$\hat{H}_{\text{lead}}=\sum_{lk}\varepsilon_{lk}\hat{d}^\dag_{lk}\hat{d}_{lk}$ 
that has a constant temperature $T$ and chemical potential $\mu_l$ ($l=\{\text{S}_1,\ \text{S}_2,\ \text{D}\}$). The interaction part of the Hamiltonian $\hat{H}_{\text{int}}=\sum_{li}\gamma_{l}\hat{d}^\dag_{l}\hat{c}_i+\text{H.c.}$ couples the reservoirs and the dots with a hopping parameter $\gamma_{l}$. The energy levels are tuned with the gate voltages present in the experiment ($U_{\text{G1}},\ U_{\text{G2}},\ U_{\text{G3}},\ U_{\text{G4}}$).
\\
The rates between the leads and the dots for incoming $(+)$ and outgoing $(-)$ electrons with respect to the dot system are given by Fermi's golden rule $\Gamma^{(+)}_{i\leftarrow l}=2\pi/\hbar|\gamma_{l}|^2 f(\mu_{l}-\epsilon_i)$ and $\Gamma^{(-)}_{l\leftarrow i}=2\pi/\hbar|\gamma_{l}|^2[1-f(\mu_{l}-\epsilon_i)]$ where $f$ is the Fermi distribution function. $\Gamma_{l}\equiv 2\pi/\hbar|\gamma_l|^2$ is smaller than the interdot coupling $\tau_{ij}$, thus we can apply the Born-Markov approximation \cite{BOOKBMS} for the interaction of the system with the leads. From the Von Neumann equation $\partial_t\varrho(t)=i/\hbar[\hat{H},\varrho(t)]$, which contains the full system time evolution, we trace over the baths degrees of freedom getting the reduce density matrix  $\rho(t)=\text{Tr}_{\text{leads}}\varrho(t)$ \cite{PhysRevB.53.1050} obtaining the master equation
\begin{align}
\partial_t\rho_i(t)=\sum_j\mathcal{L}_{ij}\rho_j(t). \label{Eq::Master}
\end{align} 
$\rho_i(t)$ is the occupation probability of the $i$-state of the system and $\mathcal{L}$ is the Liouvillian superoperator that contains all the information about the system $H_0$ and the jumping terms between the leads and the dots $\Gamma_{i\leftrightarrow l}$. As we just want to study the steady state properties of the system we solve the kernel of Eq. \eqref{Eq::Master}  to obtain the steady state occupations $\rho^{\text{SS}}=\text{Ker}[\mathcal{L}]$. 
\\
Taking the steady state occupations and the tunneling rates to/from the contacts we are able to calculate the current
\begin{align}
I&=\sum_{i,j=0}^1\rho^{\text{SS}}_{\ket{1,i,j}}\Gamma^{(-)}_{\ket{0,i,j}\leftarrow\ket{1,i,j}}-
\rho^{\text{SS}}_{\ket{0,i,j}}\Gamma^{(+)}_{\ket{1,i,j}\leftarrow\ket{0,i,j}}.
\end{align}

\end{document}